\documentclass{aa}  

\usepackage{graphicx}
\usepackage{natbib}
\usepackage[colorlinks=true,allcolors=blue,urlcolor=blue]{hyperref}
\usepackage{txfonts}

\newcommand{\Hm}{H$^-$}
\newcommand{\Hmol}{H$_2$}

\newcommand{\cmps}{\mbox{cm}^3 \mbox{s}^{-1}}

\newcommand{\colorb}{black}
\newcommand{\colorc}{black}
\newcommand{\balder}{\textcolor{\colorc}{\texttt{Balder}}}
\newcommand{\blue}{\textcolor{\colorc}{\texttt{Blue}}}
\newcommand{\multitd}{\textcolor{\colorb}{\texttt{Multi3D}}}
\newcommand{\marcs}{\textcolor{\colorb}{\texttt{MARCS}}}

\begin{document} 

\title{Revisiting the statistical equilibrium of H$^-$ in stellar atmospheres}
\subtitle{}

\author{Paul S. Barklem 
         \and
        Anish M. Amarsi
          }

\institute{Theoretical Astrophysics, Department of Physics and Astronomy, Uppsala University, Box 516, SE-751 20 Uppsala, Sweden 
           }

\date{Received 18 June 2024/ Accepted 26 July 2024}
 
\abstract{The negative hydrogen ion \Hm\ is, almost without exception, treated in local thermodynamic equilibrium (LTE) in the modelling of F, G, and K stars, where it is the dominant opacity source in the visual spectral region.  This assumption rests in practice on a study from the 1960s.  Since that work, knowledge of relevant atomic processes and theoretical calculations of stellar atmospheres and their spectra have advanced significantly, but this question has not been reexamined. We present calculations based on a slightly modified analytical model that includes H, \Hmol\, and \Hm, together with modern atomic data and a grid of 1D LTE theoretical stellar atmosphere models with stellar parameters ranging from T$_\mathrm{eff} = 4000$ to 7000~K, $\log{g} = 1$ to 5 cm/s$^2$, and [Fe/H]$=-3$ to 0.  We find direct non-LTE effects on populations in spectrum-forming regions, continua, and spectral lines of about 1--2\% in stars with higher T$_\mathrm{eff}$ and/or lower $\log g$.  Effects in models for solar parameters are smaller by a factor of 10, about 0.1--0.2\%, and are practically absent in models with lower T$_\mathrm{eff}$ and/or higher $\log g$. These departures from LTE found in our calculations originate from the radiative recombination of electrons with hydrogen to form \Hm\ exceeding photodetachment, that is, overrecombination. Modern atomic data are not a source of significant differences compared to the previous work, although detailed data for processes on \Hmol\ resolved with vibrational and rotational states provide a more complete and complex picture of the role of \Hmol\ in the equilibrium of \Hm.   In the context of modern studies of stellar spectra at the percent level, our results suggest that this question requires further attention, including a more extensive reaction network, and indirect effects due to non-LTE electron populations.}

\maketitle

%-------------------------------------------------------------------

\section{Introduction}

That the negative hydrogen ion \Hm\ contributes significantly to the opacity in the atmospheres of solar-type stars was first suggested by \cite{WildtNegativeIonsHydrogen1939}, and it is now well known that it dominates the continuous opacity in the visual and near-infrared region of F, G, and K stars, including in the Sun \citep[][Sect.~6]{ChandrasekharContinuousAbsorptionCoefficient1946}; see also, e.g. \citet[][Sect.~48]{unsold_physik_1955} and \citet[][Sect.~4-19]{aller_astrophysics_1963}.  Theoretical modelling of the atmospheres and spectra of these stars usually assumes local thermodynamic equilibrium (LTE), including for \Hm (e.g. in 1D with the \marcs{} \citep{Gustafsson2008} or \texttt{ATLAS} \citep{kurucz_atlas12_2013} codes, or in 3D with the \texttt{STAGGER} \citep{magicStaggergridGrid3D2013} or \texttt{COBOLD} \citep{freytagSimulationsStellarConvection2012} codes).

Potential departures from LTE, non-LTE effects, can take two forms for \Hm.  There is a direct effect on \Hm\ through the effects of non-local radiation on \Hm\ in competition with local collisional processes, such that the statistical equilibrium of \Hm\ differs from LTE.  There is also an indirect effect via other species.  In particular, the ionisation of electron donor elements in non-LTE may be different than that in LTE, leading to a different electron density and thus a different proportion of \Hm\ compared to a pure LTE situation.  It is also worth noting that these direct and indirect effects are not independent; for example, a changed electron density affects the collisional processes in the statistical equilibrium.  Furthermore, there would be feedback effects on the structure of the model atmosphere.

For a few late-type stars, model calculations solving the atmospheric structure self-consistently with a non-LTE treatment of a significant number of important species and their line blanketing have been performed \citep{shortAtmosphericModelsRed2003,shortNonLTELineBlanketedModel2005,shortNonLTEModelingNearUltraviolet2009} with the PHOENIX code \citep{hauschildtNumericalSolutionExpanding1999}.  \Hm\ itself is treated in LTE, that is, indirect effects are included, and no direct effect.  The total impact of non-LTE on the continuum flux for the Sun in the visual region compared to a fully LTE model is about 5\% (see Fig. 4 of \citealt{shortNonLTELineBlanketedModel2005}), although the non-LTE PHOENIX model analysed by \cite{Pereira2013} shows much smaller differences relative to LTE, where small differences in the centre-limb variations are also predicted.
While the code PHOENIX might in principle be used to produce non-LTE line-blanketed models for FGK stars, the most recent extensive grid is produced in LTE \citep{husserNewExtensiveLibrary2013}. This may reflect the computational and physical difficulties, not least in obtaining complete and accurate collision data for all relevant species, which are often not available or have only become available recently, for example for Fe \citep{wangElectronScatteringNeutral2018,barklemExcitationChargeTransfer2018}.  Thus, LTE models are used in essentially all analyses of late-type stars.

Semi-empirical modelling of the solar atmosphere such as that of \cite{VernazzaStructureSolarChromosphere1973, Vernazza1981} included a non-LTE treatment of \Hm, but these early studies focussed on the upper atmosphere and did not show significant effects on \Hm\ in the photosphere.  However, more recently,  \cite{shapiroNLTESolarIrradiance2010} found non-LTE effects of about 10\% on the \Hm\ population in the photosphere, and correspondingly large effects on the continuum flux in the visual region.  These models considered both direct and indirect effects. The large non-LTE effects on \Hm\ in the photosphere found in \cite{shapiroNLTESolarIrradiance2010} stem predominantly from the indirect effect, which is due to overionisation of metals with a low ionisation potential such as Fe, Si, and Mg (see their Sect. 4.2.1 and Fig.~8, as well as Fig. 1 of \cite{shortNonLTELineBlanketedModel2005}).

The validity of any assumption that the relative abundance of \Hm\ in cool stellar atmospheres can be predicted by LTE, that is, Saha-Boltzmann statistics, or in other words, that the direct effect is negligible, in practice rests on the study of \cite{lambert_dissociation_1968} (hereafter LP68).  This is also the basis of the statistical equilibrium model that was used in semi-empirical studies of the solar atmosphere we discussed in the previous paragraph. LP68 reduced the problem to the three most important processes involving \Hm, H, and H$_2$, enabling them to solve the statistical equilibrium problem analytically.  Their work was somewhat hampered by uncertain collision rates for the two important collision processes, namely three-body recombination to form H$_2$ from H, and the reverse process of collisional dissociation,
\begin{equation}
\mathrm{3H} \leftrightarrow \mathrm{H}_2 + \mathrm{H},
\end{equation}
and associative detachment to form H$_2$ and destroy \Hm, with the reverse process of dissociative attachment to form \Hm,
\begin{equation}
\mathrm{H} + \mathrm{H}^- \leftrightarrow \mathrm{H}_2 + e.
\end{equation}

Despite the uncertainties in the atomic data used in their study, LP68 made a convincing case that direct non-LTE effects on \Hm\ in late-type stars are small.  However, there are several reasons to revisit this question.  First, improved atomic data are available for these and other processes; the improvements are often driven by modelling hydrogen in the early Universe \cite[e.g.][]{Lepp2002,galliDawnChemistry2013} and primordial star formation \citep[e.g.][]{turkEFFECTSVARYINGTHREEBODY2010,Kreckel2010}, and in molecular clouds \citep[e.g.][]{Dalgarno2002,nesterenokCtypeShockModelling2019}.  Secondly, the current modelling of stellar spectra in 3D and non-LTE attempts to obtain accuracies at the percent level across a wide range of stellar parameters \citep[e.g.][]{AmarsiCarbonoxygeniron2019,giribaldiTITANSMetalpoorReference2021,wangDetailedChemicalCompositions2022} for use in many contexts, including large-scale high-resolution spectroscopic surveys and studies of exoplanets and their host stars \citep[e.g.][]{nissenHighprecisionStellarAbundances2018,jofreAccuracyPrecisionIndustrial2019}.  Thus, even small effects on the percent level are of interest in a modern context.  LP68 only used scaling relations for stars beyond the Sun and did not consider bright giants or supergiants (i.e. luminosity class I and II).  Finally, LP68 only considered continua and not line formation.

\cite{Litesequilibriumusingcoupled1984} later formulated a more general statistical equilibrium problem, including other species and internal states. A model like this would require a numerical solution, and even though this is possible with modern computers and numerical methods \citep[e.g.][]{Scharmer1985,rybickiAcceleratedLambdaIteration1992,arramyJacobianFreeNewtonKrylovMethod2024}, this has not yet been attempted.  It is important that these calculations are made to understand the statistical equilibrium of \Hm\ in detail, including all possible processes.  As a basis for further work, however, including understanding of any direct effects seen in more complex calculations, and since the LP68 model is still used \citep[e.g.][]{shapiroNLTESolarIrradiance2010}, we see it as an important first step to revisit the analytic model and to understand any differences compared to LP68.  This is the subject of this paper.

%--------------------------------------------------------------------
\section{Theory and model}
\label{sect:model}

\begin{figure}
\centering
\includegraphics[width=0.25\textwidth]{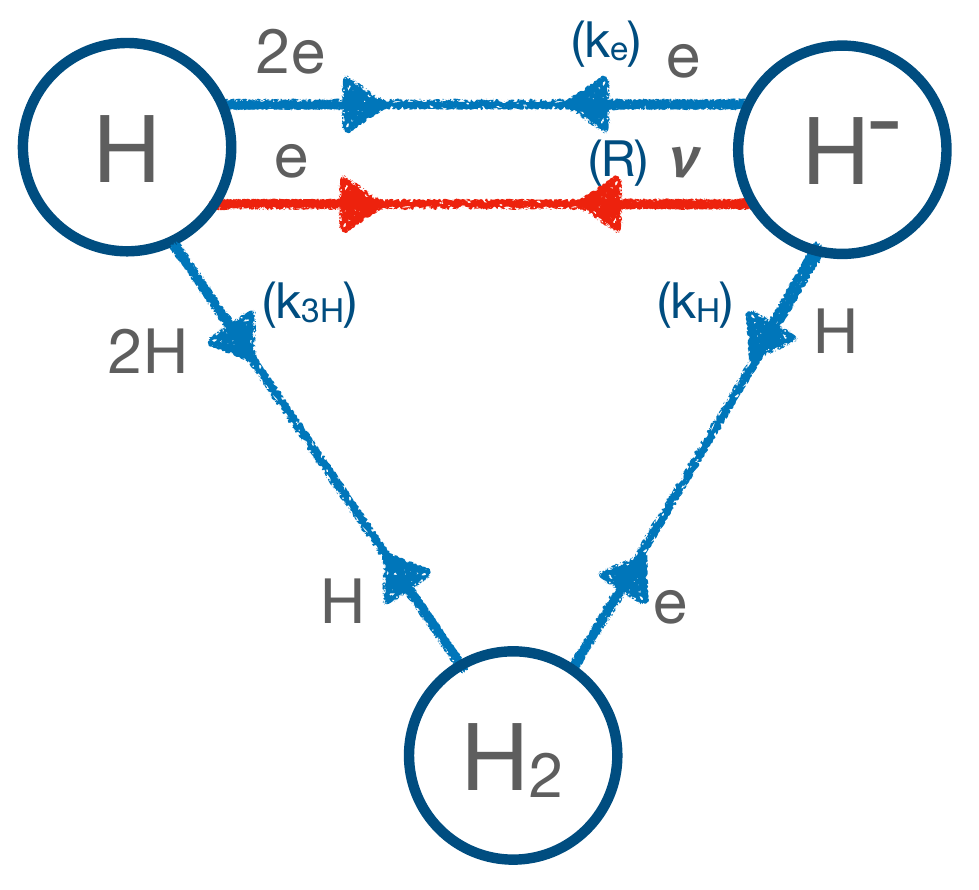}
\caption{Bubble diagram of the model and the included processes.  Collision processes are drawn in blue, and the radiative processes (involving a photon) are drawn in red, where the other particles involved in the process are shown along the relevant line. The corresponding rate coefficients for processes that appear in the model are shown in brackets: $R$ for photodetachment (see Eq.~(\ref{eq:photo}) and Sect.~\ref{sect:photo}), $k_\mathrm{H}$ for associative detachment (see Sect.~\ref{sect:ad}),  $k_\mathrm{3H}$ for three-body recombination of hydrogen (see Sect.~\ref{sect:3H}), and $k_{e}$ for electron collisional detachment (see Sect.~\ref{sect:e}). }
\label{fig:bubble}
\end{figure}

We employed an analytical model for the statistical equilibrium of \Hm\ similar to that described in \cite{VernazzaStructureSolarChromosphere1973} (pg. 619, hereafter VAL73), which is a development of the analytical model of LP68.  The model includes \Hm, \Hmol, and H, and is shown schematically in Fig.~\ref{fig:bubble}.  The difference compared to the LP68 model is that it includes the electron collisional detachment process $\mathrm{H}^- + e \leftrightarrow \mathrm{H} + 2e$ and a more precise expression for the total radiative recombination, in which stimulated emission is treated using the actual mean intensity $J_\nu$, rather than the Planck function $B_\nu$.  As in LP68 and VAL73, we assumed that \Hmol\ can be represented by a single level, and this approximation is discussed further in Sect.~\ref{sect:h2}. 

The bound-free source function for \Hm\ is given by
\begin{equation}
S_\nu = \frac{2h\nu^3}{c^2} \frac{1}{\frac{b_-}{b_1} e^{h\nu/kT} - 1},
\end{equation}
where the symbols have their usual meanings \citep[e.g.][]{ruttenRadiativeTransferStellar2003} with the subscript ``$-$'' referring to \Hm\ and subscript ``$1$'' to the ground state of hydrogen.  The departure coefficients $b$ are defined 
\begin{equation}
b_- = \frac{n_-}{n_-^\mathrm{*}} \; \mathrm{and} \; b_{1} = \frac{n_1}{n_1^\mathrm{*}}, 
\label{eq:bs}
\end{equation}
where $n$ are particle densities, and $n^\mathrm{*}$ denotes particle densities in LTE.  We used a definition for the departure coefficients, $b = n/n^*$, that is standard today \citep[e.g.][Sect.~2.6]{ruttenRadiativeTransferStellar2003}, rather than the definition used in LP68, where the relation to the modern notation, for example, is $b_-^\mathrm{LP68} = b_-/b_1$ (see Rutten's lecture notes for further details).

Writing the statistical equilibrium equations for \Hm\ and \Hmol\ including the processes shown in Fig.~\ref{fig:bubble}, which are described further below, employing the detailed balance relations for inverse processes and solving the equations, we find the departure coefficient for \Hm\ as
\begin{equation}
b_{-}  = b_1 \frac{R^\dagger + r}{R + r},
\label{eq:bm}
\end{equation}
where $R$ is the photodetachment rate per \Hm ion
\begin{equation}
R = 4\pi \int_{\nu_0}^\infty \sigma(\nu) J_\nu \frac{d\nu}{h\nu},
\label{eq:photo}
\end{equation}
with $\sigma(\nu)$ the photodetachment cross section with the frequency threshold $\nu_0$, and $R^\dagger$ is 
\begin{equation}
R^\dagger = 4\pi \int_{\nu_0}^\infty \sigma(\nu) \left( \frac{2h\nu^3}{c^2} + J_\nu \right) e^{-h\nu/kT} \frac{d\nu}{h\nu},
\end{equation}
such that the total radiative recombination rate (including spontaneous and induced recombination) is $n_{1} (n^*_{-}/n^*_{1}) R^\dagger$ \citep[e.g.][]{mihalas_stellar_1978,ruttenRadiativeTransferStellar2003,hubeny_theory_2014}.  
The quantity $r$ is defined as 
\begin{equation}
r = n_e k_e + \frac{n_{1} k_\mathrm{H}}{1 + \Omega},
\end{equation}
with
\begin{equation}
\Omega = \frac{n^*_{-} k_\mathrm{H}}{n_1^* n_1 k_\mathrm{3H}},
\end{equation}
where $k_e$ is the rate coefficient (L$^3$T$^{-1}$) for electron detachment, $k_\mathrm{H}$ is the rate coefficient (L$^3$T$^{-1}$) for associative detachment, and $k_\mathrm{3H}$ is the rate coefficient (L$^6$T$^{-1}$) for three-body recombination; these rate coefficients are discussed below.  The quantity $r$ may be interpreted as an effective total collision rate per \Hm\ ion, coupling it to H. Thus, $b_{-}$ as given by Eq.~(\ref{eq:bm}), shows the expected behaviour that when $b_1 = 1$, and collisions dominate such that $r\gg R$ and $r\gg R^\dagger$, then $b_{-} \rightarrow 1$.  When $b_1 = 1$ and the radiative process dominates such that  $r\ll R$ and $r\ll R^\dagger$, then $b_{-} \rightarrow R^\dagger/R$.  In LTE ($J_\nu = B_\nu$), again denoting LTE values with an asterisk superscript, we have
\begin{equation}
(R^\dagger)^* = 4\pi \int_{\nu_0}^\infty \sigma(\nu) B_\nu \frac{d\nu}{h\nu} = R^*,
\end{equation}
and $b_{-} = b_{1}$.

We note that our results are very similar to but not precisely the same as those found in VAL73; their equations seem only to be strictly valid if $b_1=1$. In our derivation we distinguished between $n_1$, the density for ground-state hydrogen atoms, and $n_H$, the total hydrogen atom density, although in practice at the temperatures of interest, they will often be very similar.

Finally, the departure coefficient for \Hmol, denoted $b_{\mathrm{H}_2}$, can be found to be
\begin{equation}
b_{\mathrm{H}_2} = \frac{n_{\mathrm{H}_2}}{n^*_{\mathrm{H}_2}} = \frac{1 + (b_{-}/b_1) \Omega}{1 + \Omega}.
\end{equation}
When $\Omega = 0$, and therefore, no coupling of \Hmol\ to \Hm, then $b_{\mathrm{H}_2}=1$.  In practice, $\Omega > 0$, and it can be seen that $b_{\mathrm{H}_2}$ is always closer to 1 than $b_{-}/b_1$ is to 1, and that if $b_{-}/b_1 > 1 $, then $b_{\mathrm{H}_2} > 1$, and if $b_{-}/b_1 < 1 $, then $b_{\mathrm{H}_2} < 1$.
 
 \subsection{Processes and rates}
 
We now discuss the processes we employed in the model and the data we used for the rates in each case.  In Table~\ref{tab:rates}, we compare the rate coefficients for collision processes we adopted with those used by LP68.  Even though the data used in LP68 were uncertain, the data they used for the important cases of $k_\mathrm{H}$ and $k_\mathrm{3H}$ agree reasonably well with modern data at the temperatures of interest.  Electron collisional detachment $k_e$ was discussed by LP68, but was not included in the calculations, and Table~\ref{tab:rates} shows that modern data suggest that this process is even less efficient than the estimate discussed in LP68.

\begin{table*}
\caption{Comparison of rate coefficients for collision processes at temperatures of interest.}  
\label{tab:rates}  
\centering
\small
\begin{tabular}{c |ccc |ccc |ccc}
\hline\hline
T & \multicolumn{3}{c|}{$k_\mathrm{H}$} & \multicolumn{3}{c|}{$k_\mathrm{3H}$} & \multicolumn{3}{c}{$k_e$} \\
  & this work & LP68 & ratio & this work & LP68 & ratio & this work & LP68 & ratio \\
{} [K] & [cm$^3$ s$^{-1}$] & [cm$^3$ s$^{-1}$] & & [cm$^6$ s$^{-1}$] & [cm$^6$ s$^{-1}$] &  & [cm$^3$ s$^{-1}$] & [cm$^3$ s$^{-1}$] & \\  
\hline       
     & &&& &&& && \\             
3000 & $1.75 \times 10^{-9}$ & $1.26 \times 10^{-9}$ & $1.39$ & $1.18 \times 10^{-32}$ & $1.68 \times 10^{-32}$ & $0.70$ & $2.06 \times 10^{-12}$ & $9.18 \times 10^{-10}$ & $0.0022$ \\               
5000 & $1.49 \times 10^{-9}$ & $1.26 \times 10^{-9}$ & $1.18$ & $9.96 \times 10^{-33}$ & $1.01 \times 10^{-32}$ & $0.99$ & $4.85 \times 10^{-11}$ & $1.98 \times 10^{-9}$ & $0.025$ \\             
7000 & $1.29 \times 10^{-9}$ & $1.26 \times 10^{-9}$ & $1.02$ & $8.95 \times 10^{-33}$ & $7.20 \times 10^{-33}$ & $1.24$ & $2.00 \times 10^{-10}$ & $3.27 \times 10^{-9}$ & $0.061$ \\ 
\hline
\end{tabular}
\tablefoot{The data adopted in this work are described in the text, and for $k_H$, they are taken from \cite{Kreckel2010}, for $k_{3H}$ from \cite{ForreyRateFormationHydrogen2013}, and for $k_e$ based on \cite{AndersenElectronImpactDetachmentNearThreshold1995} and \cite{Vejby-ChristensenElectronimpactdetachmentnegative1996}.  The data adopted by LP68 are for $k_H$ based on \cite{dalgarnoAssociativeDetachment1967}, for $k_{3H}$ based on a semi-empirical expression (their Eq. (11)), and for $k_e$ from \cite{thomasPhysicsSolarChromosphere1961}.}
\end{table*}

\subsubsection{Photodetachment and recombination: $\mathrm{H}^- + h\nu \leftrightarrow \mathrm{H} + e$ \hspace{1cm}  [$\sigma(\nu)$]}
\label{sect:photo}

The rate for photodetachment given by Eq.~(\ref{eq:photo}) requires the cross section $\sigma(\nu)$.  The most commonly used data are from \cite{Wishartboundfreephotodetachmentcrosssection1979} and cover 16300 \AA\  (threshold, 0.751 eV) to 1250 \AA\ (almost 10 eV).  \cite{McLaughlinHphotodetachmentradiativeattachment2017} have presented new calculations including resonances at high energy ($>10$ eV) in the far-UV (first resonance at roughly 1120 \AA).  The differences compared to the  \cite{Wishartboundfreephotodetachmentcrosssection1979} data are very small in the visual and UV (photon energies up to 7 eV, roughly redward of 1700 \AA; \citealt[][see Fig. 1]{McLaughlinHphotodetachmentradiativeattachment2017}), and there are numerous theoretical and experimental results that agree well in the region up to 3 eV \citep[][see Fig. 2]{McLaughlinHphotodetachmentradiativeattachment2017}.  At higher energies, the results depart, not least due to the resonances.  For the stars of interest, the radiative rate is dominated by the contribution from the visual, and we used the cross sections from \cite{Wishartboundfreephotodetachmentcrosssection1979}. 

\subsubsection{Associative detachment: $\mathrm{H}^- + \mathrm{H} \leftrightarrow \mathrm{H}_2 + e$ \hspace{1cm} [$k_\mathrm{H}$]}
\label{sect:ad}

The rate for associative detachment is $n_- n_1 k_\mathrm{H}$.  We employed the total rate coefficient from \cite{Kreckel2010}, the fit given in their Table~S1, which is based on the theoretical method developed in \cite{CizekNucleardynamicscollision1998}, and shown by \cite{Kreckel2010} to agree very well with their results from a merged-beams experiment.  State-resolved data for the reverse process of dissociative attachment were also presented in \cite{launayReversibleYieldsH21991}.  These give a lower total associative detachment rate than the data used here (see Fig.~S3 of \cite{Kreckel2010}).  

LP68 employed a temperature-independent estimate of this rate coefficient based on calculations by \cite{dalgarnoAssociativeDetachment1967}.  As listed in Table~\ref{tab:rates}, this value is within 40\% of the data adopted in this work, although modern data suggest a significant temperature dependence.

\subsubsection{Three-body recombination collisions: $3\mathrm{H} \leftrightarrow \mathrm{H}_2 + \mathrm{H}$ \hspace{1cm} [$k_\mathrm{3H}$]}
\label{sect:3H}

The rate for three-body recombination of hydrogen forming \Hmol\ is $ (n_1)^3 k_\mathrm{3H}$.  \cite{ForreyRateFormationHydrogen2013} have performed calculations based on a Sturmian theory of three-body recombination \citep{ForreySturmiantheorythreebody2013}, and reported a fit to the rate coefficients valid over the range 300~K to 10000~K,
\begin{equation}
k_\mathrm{3H} = 6 \times 10^{-32} T^{-1/4} + 2 \times 10^{-31} T^{-1/2} \, [\mathrm{cm}^6/\mathrm{s}],
\end{equation}
for $T$ in K, and we used this rate coefficient in our modelling.  This is estimated to be accurate to roughly a factor of two.   This result gives a much flatter temperature dependence than previous results of \cite{JacobsKineticsHydrogenHalides1967}, \cite{PallaPrimordialstarformation1983}, \cite{AbelFormationFirstStar2002}, and \cite{FlowerThreebodyrecombinationhydrogen2007}.  In addition, it is much lower than that from \citeauthor{FlowerThreebodyrecombinationhydrogen2007} (by roughly a factor 2--3 at $T \sim 5000$--7000 K) and significantly higher than that from \citeauthor{AbelFormationFirstStar2002} (slightly more than a factor of 10 at T$ \sim 5000$--7000 K); see Fig.~1 of \cite{ForreyRateFormationHydrogen2013}.  State-resolved data for the reverse process of three-body collisional dissociation have also been presented in \cite{bossionRovibrationalExcitationH22018}.  

LP68 employed a semi-empirical estimate of this rate coefficient based on experiments with a large scatter.  As listed in Table~\ref{tab:rates}, their estimate generally agrees reasonably well with the data adopted in this work, although modern theoretical data suggest a stronger temperature dependence.

\subsubsection{Electron collisional detachment and recombination: $\mathrm{H}^- + e \leftrightarrow \mathrm{H} + 2e$ \hspace{1cm} [$k_e$]}
\label{sect:e}

The rate for collisional detachment of \Hm\ by electron collisions is $n_- n_e k_\mathrm{e}$.  Various early measurements of the cross section for this process were made at energies well above the threshold \citep[e.g.][]{TisoneDetachmentElectronsElectron1966, DanceMeasurementCrossSectionDetachment1967, TisoneDetachmentElectronsNegative1968, PeartElectrondetachmentions1970}.  The first measurement near the threshold was by \cite{Waltonmeasurementcrosssections1971}.  Later storage ring experiments probed the low-energy regime in detail, first for D$^-$ \citep{AndersenElectronImpactDetachmentNearThreshold1995, Vejby-ChristensenElectronimpactdetachmentnegative1996}, and later for \Hm\ \citep{FritioffSingledoubledetachment2004}.

The experiments at low energy show a very small, possibly zero, cross section at energies just above the electron affinity of \Hm, about 0.75 eV, until an observed threshold at about 2 eV, with the cross section then increasing rapidly to roughly 100 $a_0^2$ at about 20 eV.  This general behaviour is reasonably explained with a classical over-the-barrier model. The negligible cross section between 0.75 eV and roughly 2 eV due to the Coulomb repulsion means that the low-energy electron is unable to approach the \Hm\ ion sufficiently closely for the detachment process to occur.

For solar temperature stellar atmospheres, average collision energies are $kT\sim 0.2$ to 0.6 eV, and they are therefore below threshold for this process. We must therefore expect this process to be inefficient.  A reasonable estimate for the cross section $\sigma$ as a function of collision energy $E$ can be made from the classical model \citep{AndersenElectronImpactDetachmentNearThreshold1995, Vejby-ChristensenElectronimpactdetachmentnegative1996},
\begin{equation}
\sigma(E) = p \pi R^2 \, \mathrm{max}[0, (1-U_c/E)],
\end{equation}
where $U_c = 27.2/(R/a_0)$~eV, and the best-fit parameters are $p=0.2$ and $R=14.5 a_0$.
Using this cross section, we find that the rate coefficient integrated over a Maxwellian velocity distribution can be well represented over the range $T=40$--$10^5$ K by
\begin{equation}
k_e  =   \langle \sigma \varv \rangle  = a_1 (T/300)^{a_2} \exp{(-a_3/T)}  
\end{equation}
with parameters $a_1 = 9.28372 \times 10^{-10} \, \cmps$, $a_2 = 0.5$, and $a_3 = 21787.8$~K.

LP68 considered an estimate of this rate coefficient from \cite{thomasPhysicsSolarChromosphere1961} based on cross-section calculations by \cite{geltmanElectronDetachmentNegative1960}, and concluded that this process was likely to be unimportant.  As listed in Table~\ref{tab:rates}, the rate coefficients adopted in this work are lower by roughly two to three orders of magnitude at the temperatures of interest.  

\subsection{Treatment of \Hmol}
\label{sect:h2}

LP68, VAL73, and our model all assume that the internal structure of \Hmol, specifically, the vibrational and rotational levels of the ground electronic state, does not need to be considered.  Thus, any departure from equilibrium for \Hmol\ is captured in a single overall population $n_{\mathrm{H}_2}$ and departure coefficient $b_{\mathrm{H}_2}$.  

The basis of this assumption is that \Hmol\ is likely to be in, or very close to, internal equilibrium, since the dense system of vibrational levels with quantum numbers $\varv$ and rotational levels with quantum numbers $J$ are expected to be strongly coupled through collisions.  LP68 discussed this assumption at length, and presented a simple two-level model of \Hmol\, based on the understanding at the time of which vibrational states are involved in the three-body recombination and associative detachment processes. The discussion was hindered by the fact that no theoretical or laboratory estimates of de-excitation of vibrational levels by hydrogen atoms were available.  Now, theoretical state-resolved data are available for hydrogen collisions on \Hmol\ \citep{bossionRovibrationalExcitationH22018}, including excitation and de-excitation, and three-body collisional dissociation (reverse process described in Sect.~\ref{sect:3H}).  We also have state-resolved data for associative detachment \citep{launayReversibleYieldsH21991}.  Based on these data, we extract a somewhat different and more complex picture of what happens in \Hmol\ in connection to \Hm\ formation than found by LP68, which we now discuss.

\begin{figure}
\centering
\includegraphics[trim=270 110 280 125,clip,width=0.42\textwidth]{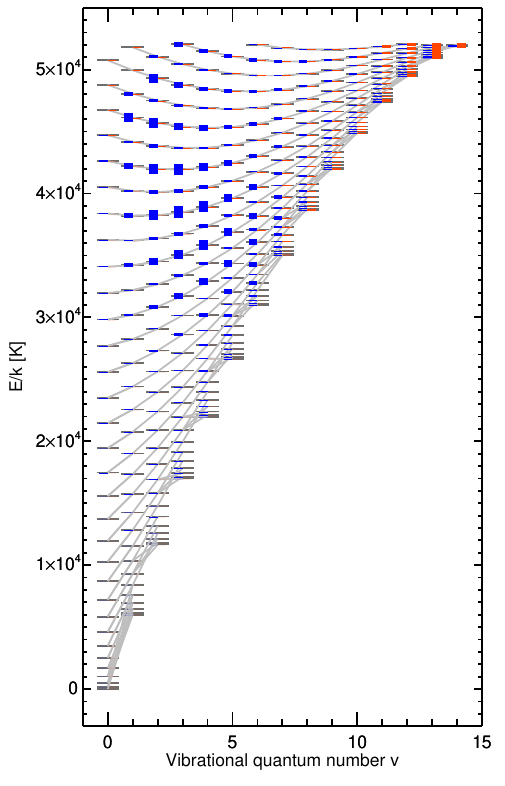}
\caption{Graphical representation of the efficiency of collision processes on \Hmol\ states at 5000~K, shown in terms of state energy on the y-axis and vibrational quantum number $\varv$ on the x-axis.  The left (blue) vertical bar for each state has a height proportional to the associative detachment rate coefficient ($\mathrm{H}^- + \mathrm{H} \rightarrow \mathrm{H}_2(\varv,J) + e$) from \cite{launayReversibleYieldsH21991}. The right (orange-red) vertical bar for each state has a height proportional to the collisional dissociation rate coefficient ($\mathrm{H}_2(\varv,J) + \mathrm{H} \rightarrow 3\mathrm{H}$) from \cite{bossionRovibrationalExcitationH22018}.  The length scales are arbitrarily chosen in each case so that the relative proportions can be seen.  For each state $(\varv,J)$, the de-excitation (most cases) or excitation transition ($\mathrm{H}_2(\varv,J) + \mathrm{H} \rightarrow \mathrm{H}_2(\varv',J') + \mathrm{H}$) with the highest rate coefficient $k$ is shown with a grey line; data from \cite{bossionRovibrationalExcitationH22018}.  }\label{fig:h2_proc} 
\end{figure}

Based on energy considerations, LP68 expected \Hmol\ to be preferentially formed in or destroyed from high vibrational states $\varv\ge 8$, the highest fully bound vibrational states having $\varv=14$ \citep{komasaQuantumElectrodynamicsEffects2011}. This has since been given support from calculations \citep{OrelNascentvibrationalrotational1987, EspositoSelectiveVibrationalPumping2009, bossionRovibrationalExcitationH22018, YuenJahnTellereffectthreebody2020}.  Figure~\ref{fig:h2_proc} shows data from \cite{bossionRovibrationalExcitationH22018} at 5000~K, and it shows that capture occurs most efficiently into states with high $\varv$ and low $J$, which are all clustered in energy close to the dissociation limit.

Furthermore, LP68 expected associative detachment to preferentially form $\varv\approx 8$, based on examination of the potential curves for H$_2$ and H$_2^-$ in \cite{ChenAssociativeDetachmentH21968}. The crossing between the potential curves lies approximately 3.7 eV above the bottom of the H$_2$ potential well, and the ($\varv$,$J$) = (0,0) state of \Hmol\ lies approximately 0.27 eV above the bottom of the well, and thus the crossing could be expected to populate states of H$_2$ with excitation about 3.4 eV above (0,0), that is, roughly $E/k\approx 40000$~K.  This led LP68 to expect the $\varv\approx 8$ state to be most strongly populated.  The modern calculations change this picture.  Extensive ro-vibrationally resolved calculations were performed by \cite{launayReversibleYieldsH21991}, and the data are shown at 5000~K in Fig.~\ref{fig:h2_proc}. The state with the highest rate for population is (3,19), which lies at roughly this level of excitation.  The production in a wide spread of $\varv$ and $J$ states, and thus energy, is also significant, however.  For example, there is significant capture into (2,27) and (3,27), with (3,27) lying very close to the dissociation limit, and thus similar in energy to states with high three-body dissociation rates.  In general, Fig.~\ref{fig:h2_proc} shows that at the temperatures of interest, associative detachment is now expected to form a wide range of states with $\varv\approx 1-7$ in rotational states $J\approx7-27$, and energies spanning from roughly 2.6 eV ($\sim$30000~K) to the dissociation limit, $\sim 4.48$~eV ($\sim$52000~K). Very similar results were reported by \cite{BieniekAssociativedetachmentcollisions1979}, but the temperature grid was much more sparse.

As regards coupling between rotational and vibrational states due to collisions with hydrogen, calculations by \cite{LiqueRevisitedstudyrovibrational2015} and by \cite{bossionRovibrationalExcitationH22018} showed that rotational de-excitation (or in some cases, excitation) rate coefficients for transitions to the nearest-lying state with $\Delta J = \pm 2$ are about $10^{-10}\, \cmps$ at temperatures of interest around 5000~K; this transition is favoured as it conserves nuclear spin, that is, ortho$\rightarrow$ortho or para$\rightarrow$para.  \cite{bossionRovibrationalExcitationH22018} also showed that vibrational transitions with $\Delta \varv=-1$ are also efficient, for example (5,5) to (4,7) has a rate coefficient of about $8 \times 10^{-11}\, \cmps$ at around 5000~K.  Figure~\ref{fig:h2_proc} also shows for each state the transition with the highest rate coefficient in the data of \cite{bossionRovibrationalExcitationH22018}, and this behaviour can be seen.  

In summary, modern data suggest that the main collisional pathway for the formation of \Hm\ is formation of \Hmol\ in states with high $\varv$ and low $J$ in three-body recombination collisions,
\begin{equation}
3\mathrm{H} \rightarrow \mathrm{H}_2 (\mathrm{high}\,\varv, \mathrm{low}\,J) + \mathrm{H},
\end{equation}
followed by collisional de-excitation (or excitation) by hydrogen to states with lower $\varv$ and larger $J$ via $\Delta J = \pm 2$ and $\Delta \varv=-1$, that is,
\begin{equation}
\mathrm{H}_2 (\mathrm{high}\,\varv, \mathrm{low}\,J) + \mathrm{H} \rightarrow \mathrm{H}_2 (\mathrm{low}\,\varv, \mathrm{high}\,J) + \mathrm{H}.
\end{equation}
This is followed by dissociative attachment from a large set of vibrational and rotational states with a wide span in energy, namely
\begin{equation}
\mathrm{H}_2 (\mathrm{low}\,\varv, \mathrm{high}\,J) + e \rightarrow  \mathrm{H}^- +  \mathrm{H}, 
\end{equation}
in total turning a hydrogen atom and an electron into a \Hm\ ion.

Given this more complex picture, we do not immediately see a way to test the assumption of neglecting the internal structure of \Hmol\ other than detailed calculations.  Simplification of the system would involve approximations that are difficult to justify in that they must involve assuming at least partial internal equilibrium, which is what we would be attempting to test.  We may simply note that LP68 concluded that if the rate coefficient for de-excitation of a collapsed level populated by three-body recombination collisions were $k > 10^{-10}\, \cmps$, then this approximation would be valid, within their model, and the individual rates are usually of this magnitude.  However, the picture seems more complicated as there are myriad possible internal paths through \Hmol\ between H and \Hm.

%--------------------------------------------------------------------
\section{Calculations}
\label{sect:calc}

Using this model for the statistical equilibrium, we have calculated departures from LTE for \Hm\ and attempted to quantify the impact of the departures on stellar spectra.  This was done via post-processing of 1D LTE model atmospheres. That is, we only studied the first-order impact on the populations, synthetic spectral lines, and synthetic continua; the feedback on the model atmosphere itself was not taken into account (the so-called restricted non-LTE problem;
e.g.~\citealt{HummerFormationSpectralLines1971}).  It should also be emphasised that since LTE model atmospheres were used, the population of neutral hydrogen in the ground state was assumed to be in LTE, $b_1=1$; this assumption in the photosphere is strongly supported by calculations (e.g. Fig. 4 of \cite{Barklem2007}). 

We selected a subset of model atmospheres, covering T$_\mathrm{eff} = [4000, 5000, 6000, 7000]$~K, $\log{g} = [1,2,3,4,5]$ cm/s$^2$, and [Fe/H]$=[-3,-2,-1,0]$, drawn from a grid of standard 1D LTE \marcs{} model atmospheres \citep{Gustafsson2008}.  The models of FGK-type stars used here are the same as those in, for example, \citet{amarsiGALAHSurveyNonLTE2020}, and full details can be found therein.  We note that plane-parallel models, computed
with microturbulence of $1\,\mathrm{km\,s^{-1}}$, were used for stars with
$\log{g}\geq4.0$, while spherical models of solar mass, computed with
microturbulence of $2\,\mathrm{km\,s^{-1}}$, were used otherwise.  Models with 
$\log{g} = 1$ for T$_\mathrm{eff} = 6000$ and 7000~K are not covered, and there is no model with T$_\mathrm{eff}/\log g/ [\mathrm{Fe}/\mathrm{H}]=7000/3.0/-$1 in this grid, leaving a total of 71 models. 
We also note
that they have a standard composition: namely, solar-scaled models based on
the solar abundance mixture of \citealt{GrevesseSolarChemicalComposition2007}, with enhancements
to $\alpha$-elements for metal-poor stars of
$\mathrm{[\alpha/Fe]}=-0.4\times\mathrm{[Fe/H]}$ for
$-1.0\leq\mathrm{[Fe/H]}\leq 0.0$ and $\mathrm{[\alpha/Fe]}=+0.4$ for
$\mathrm{[Fe/H]}\leq-1.0$.

The radiative transfer post-processing calculations were carried out using
\balder{} \citep{AmarsiEffectivetemperaturedeterminations2018,Amarsi3DnonLTEiron2022}, a modified version of
\multitd{} \citep{BotnenMulti3D3DNonLTE1999,Leenaarts2009}.  Background
opacities were determined using \blue{} \citep{Zhou3DStaggermodel2023a}.  Two sets
of calculations were carried out: the first calculation was to determine the angle-averaged
intensity $J_{\nu}$ as a function of depth, which was then used to compute
departure coefficients for $\mathrm{H^{-}}$.  In a second step, these departure
coefficients were read back into \balder{} and used in the calculation of
synthetic spectra.

For the first set of calculations (to determine $J_{\nu}$), background line
opacities were pre-computed with \blue{} on a grid from $91\,\mathrm{nm}$ to
$1644\,\mathrm{nm}$ with a uniform spacing in $\log\lambda$ such that
$R=\lambda/\Delta\lambda=50\,000$ (corresponding to $144\,703$ wavelength
points).  The angle-averaged intensity $J_{\nu}$ was then computed with
\balder{}, with Rayleigh scattering in the red wing of Lyman-$\alpha$ included
and other processes treated in pure absorption.  The calculated $J_{\nu}$ were confirmed to agree well with those given by the \marcs{} code.  Given $J_{\nu}$, departure coefficients were then computed following Sect.~\ref{sect:model}.

For the second set of calculations (to determine synthetic spectra), the
departure coefficients $b_{-}$ determined in the previous step were read back
into \balder{}.  Specifically, the contribution to the true $\mathrm{H^{-}}$
bound-free opacity was modified from $n^*_{\mathrm{H^{-}}}\sigma_{\mathrm{H^{-}}}$
to $b_{-}\times n^*_{\mathrm{H^{-}}}\sigma_{\mathrm{H^{-}}}$.  The stimulated
emission and the emissivity were not corrected (i.e.~atomic hydrogen was assumed
to be in LTE).  With this modification, we carried out a spectrum synthesis of the visual region, roughly 400 to 800 nm, including 
$142$ \ion{Fe}{II} lines following \citet{AmarsiCarbonoxygeniron2019}, with the atomic
data originating from \citet{MelendezBothaccurateprecise2009}.  The choice of these lines
was arbitrary: They are convenient because they are spread across the optical,
and they apparently show no significant deviations from LTE in 1D or 3D for much
of the parameter space \citep{Amarsi3DnonLTEiron2022}.  The background line opacity
(used in the calculation of $J_{\nu}$) was excluded from the final spectrum
synthesis for clarity.  The theoretical continuum was computed in the same
fashion, but without any \ion{Fe}{II} line opacity.

\section{Results}

We now present the results for the departure coefficients and spectra.  The spectrum synthesis calculations described above allowed us to study the effects on continua and spectral lines in the visual across our grid of models, and thus, to extract some general tendencies and behaviours. 

\subsection{Departure coefficients}
 
\begin{figure*}[!h]
\centering
\includegraphics[width=0.75\textwidth]{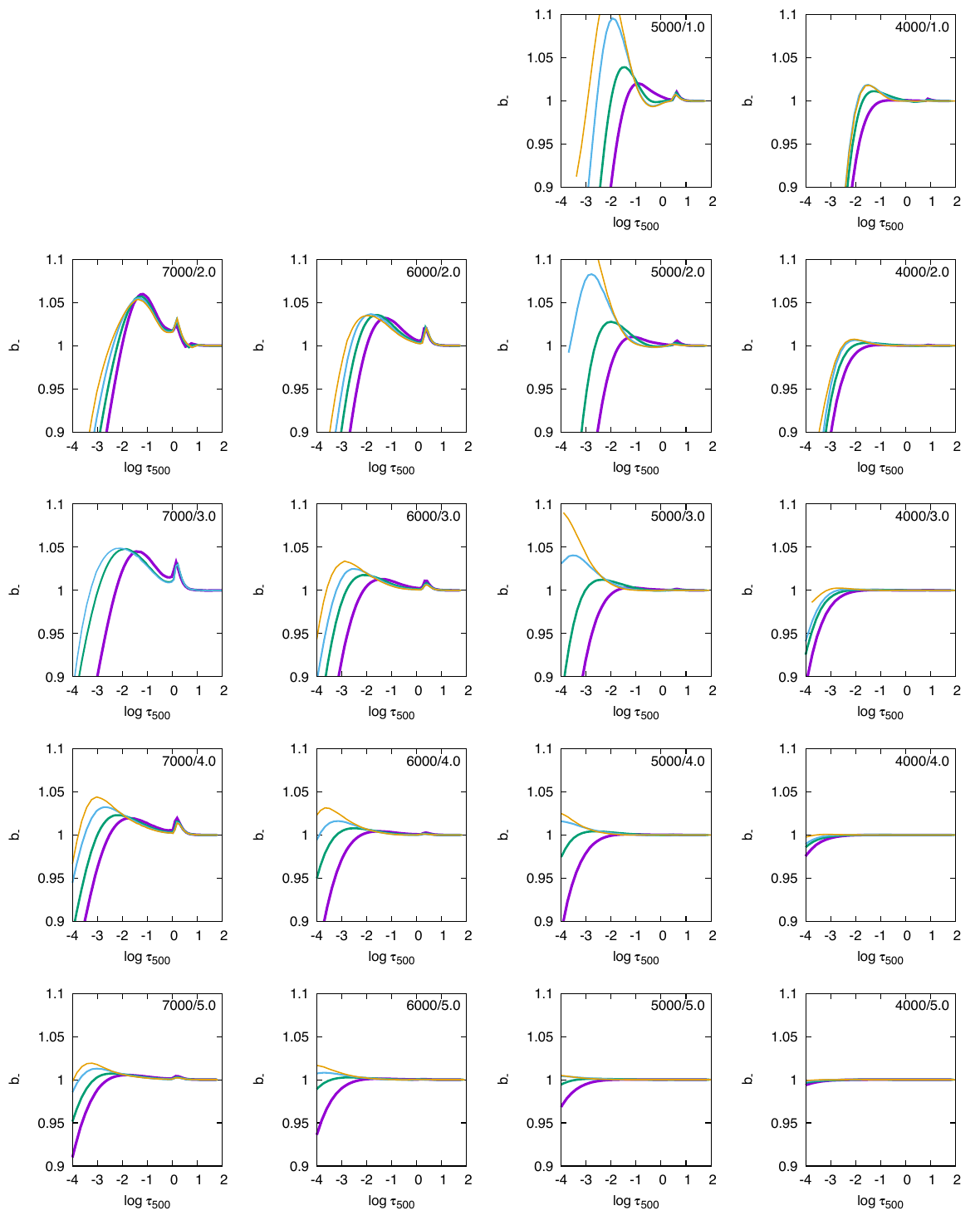}
\caption{Results for departure coefficients $b_-$ with the logarithm of the optical depth at 500 nm, $\log \tau_{500}$, across a grid of \marcs{} model atmospheres.  In the upper right corner of each panel $T_\mathrm{eff}/\log g$ is given, and each panel shows plots for different metallicity $[\mathrm{Fe}/\mathrm{H}] = $ 0, $-1$, $-2$, and $-3$.  The thickest (purple) line shows 0, and the thinnest (orange) line shows $-3$. This grid does not include a model with $T_\mathrm{eff}/\log g / [\mathrm{Fe}/\mathrm{H}]=7000/3.0/ -1$. }\label{fig:grid}
\end{figure*}

The results for $b_-$ across our grid are shown in Fig.~\ref{fig:grid}.  We focused our attention on the region around optical depth unity ($\log \tau_{500} = 0$) where the visual continuum forms, roughly $-1 \ga \log \tau_{500} \ga 0.5$ \citep[e.g.][Chapter 9]{grayObservationAnalysisStellar2022}. While the departure coefficients tend to unity in the very deepest layers, models with high T$_\mathrm{eff}$ and/or low $\log g$, show departures from LTE populations with $b_- > 1$ in this region.  Generally, the departures consist of a sharp peak just below optical depth unity and a broader peak higher in the atmosphere.  The behaviour at these depths is not very sensitive to metallicity, but there is a tendency for departures to be slightly smaller at lower metallicity. 

\begin{figure*}[!h]
%\centering
%\includegraphics[width=0.49\textwidth]{figs/hminus_conv.pdf}
%\includegraphics[width=0.49\textwidth]{figs/hminus_conv2.pdf}
\sidecaption
\includegraphics[trim=10 60 70 40,clip,width=6cm]{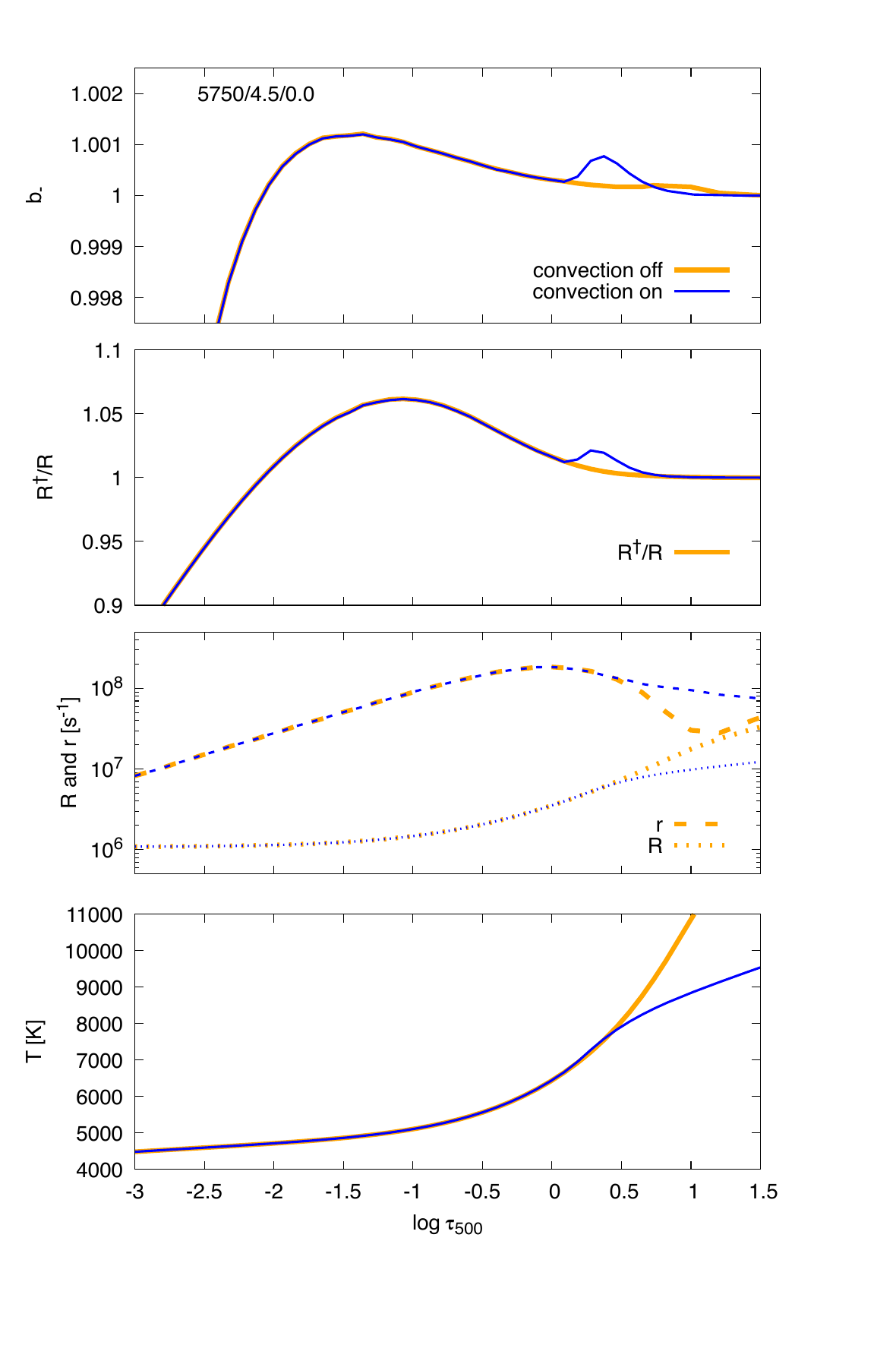}
\includegraphics[trim=10 60 70 40,clip,width=6cm]{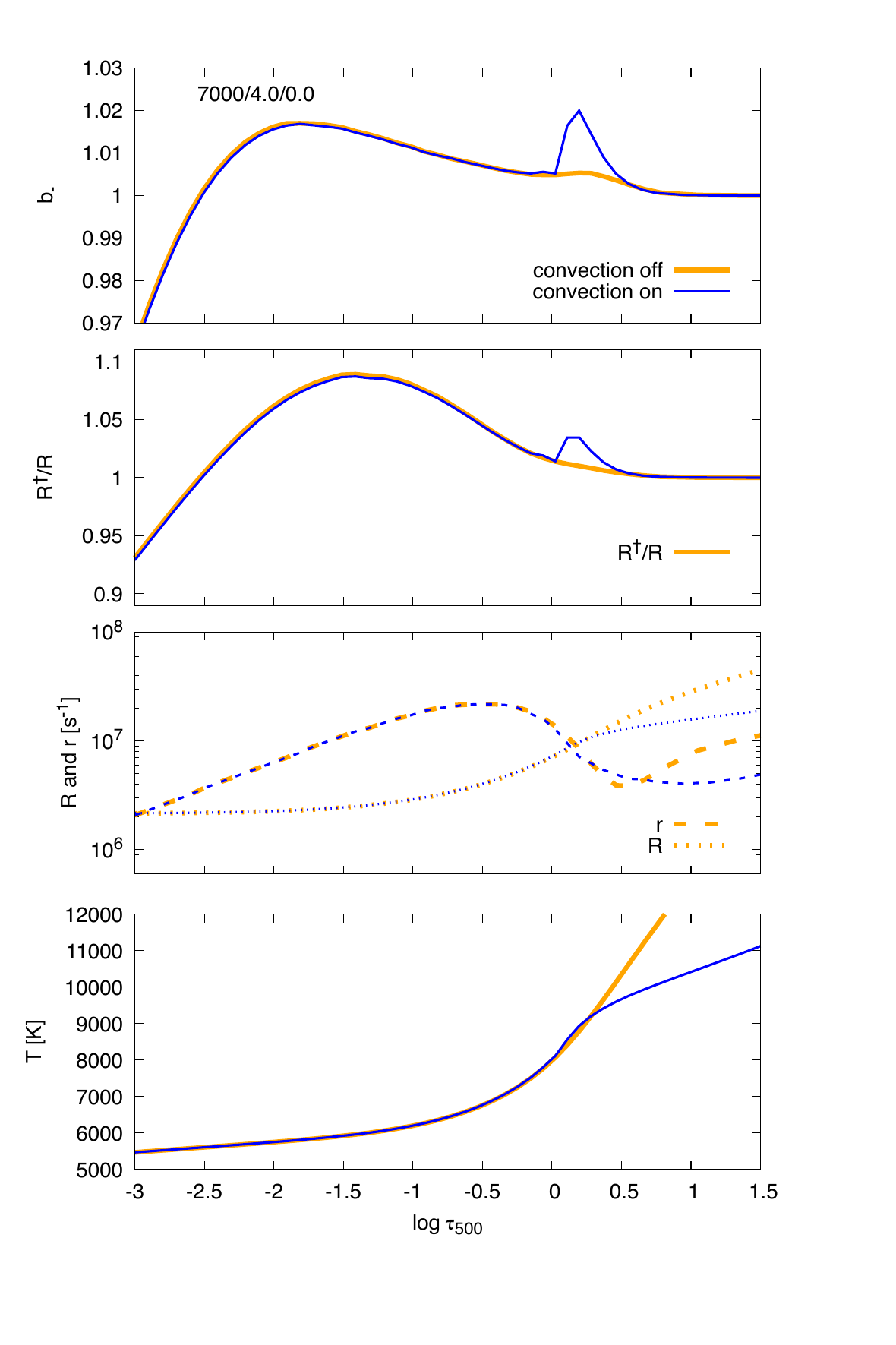}
\caption{Non-LTE mechanism for a model at roughly solar parameters (T$_\mathrm{eff} = 5750$~K, $\log g = 4.5$~cm$/$s$^2$, [Fe/H]$=0$) in the left panels and for a hotter star (T$_\mathrm{eff} = 7000$~K, $\log g = 4.0$~cm$/$s$^2$, [Fe/H]$=0$) showing larger departures from LTE in the right panels.  In each case, the upper panel shows the run of the departure coefficient $b_-$ with the logarithm of the optical depth at 500 nm, $\log \tau_{500}$.  The second panel shows the ratio $R^\dagger/R$.  The third panel shows the photodetachment rate per \Hm\ ion $R$ and effective collisional rate per \Hm\ ion $r$. The lowest panel shows the temperature $T$.  In each case, models with standard convection for \marcs{} models (mixing length parameter $\alpha=1.5$, thinner blue lines) and without convection ($\alpha=0$, thicker orange lines) are shown.}\label{fig:conv}
\end{figure*}

The reason for these departures can be seen from examination of other quantities shown for two examples in Fig.~\ref{fig:conv}.  In particular, 
the behaviour (but not the magnitude) of the departure coefficient follows $R^\dagger / R$, which is greater than unity in this region.   This indicates that radiative recombination of electrons with hydrogen to form \Hm\ at the local temperature exceeds the photodetachment resulting from the non-local radiation field, the mechanism known as overrecombination \citep{brulsFormationHelioseismologyLines1992}.  This is a result of the fact that in this part of the atmosphere, $J_\nu < B_\nu$ at infrared wavelengths, combined with the photodetachment cross section peaking at around 850 nm in the infrared, leading to $R^\dagger > R$.    
  
We show the results for models with a standard treatment of convection, mixing length efficiency parameter $\alpha=1.5$, and also without convection, $\alpha=0$.  The sharp peak below optical depth unity can be attributed to the significant flattening in temperature gradient deep in the atmosphere due to convection.  We note that even when convection is turned off completely, a small residual bump persists near this region that stems from a decline in the effective collision rate $r$, which is caused by H becoming ionised deeper in the atmosphere.    

Finally, we note that these results are not very sensitive to the chosen atomic data within reasonable limits.  For example, tests using the data from LP68 found comparable results.

\subsection{Spectrum: Continuum and spectral lines}

\begin{figure}
\centering
\includegraphics[trim=50 80 100 80,clip,width=0.49\textwidth]{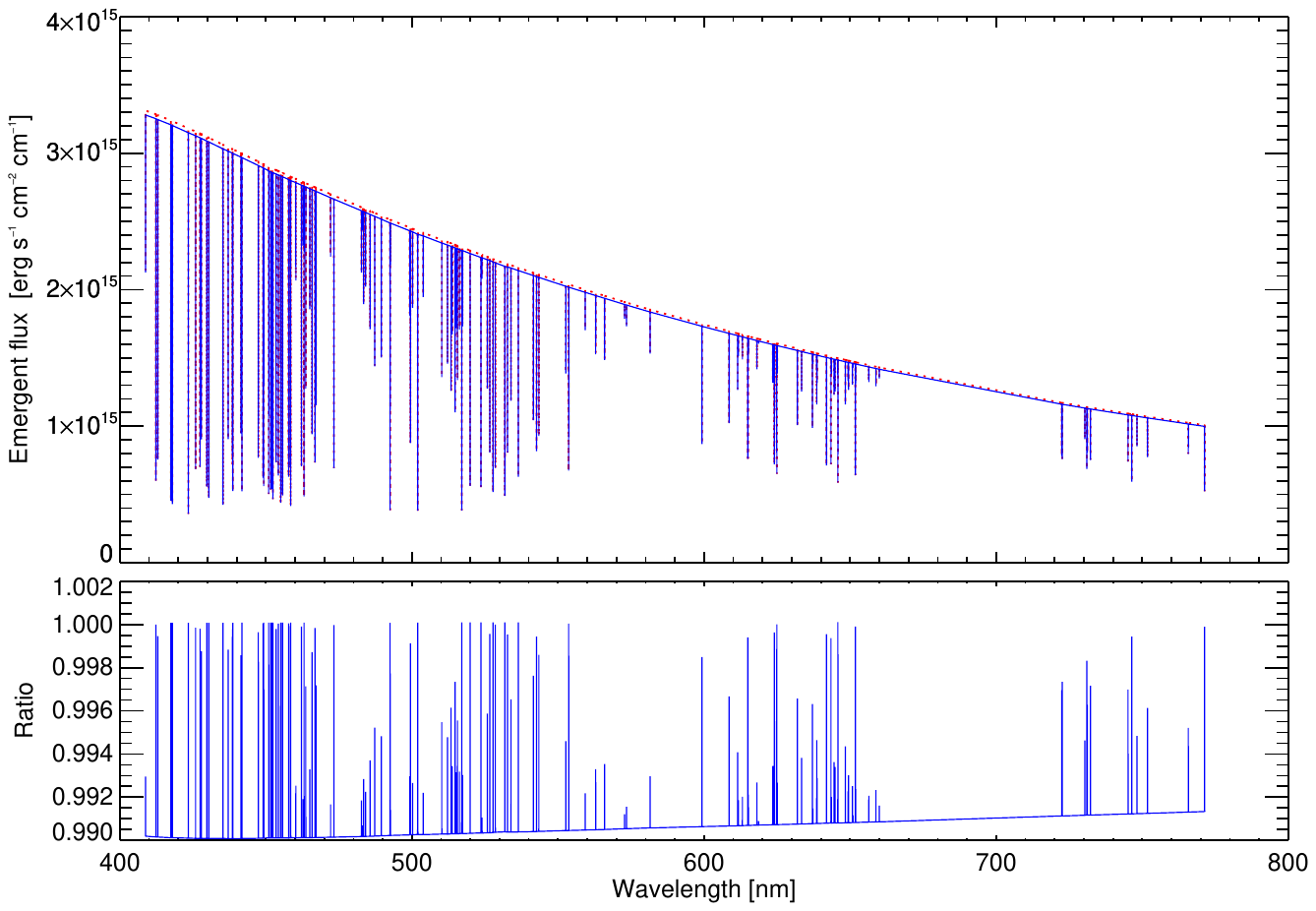}
\caption{Effect on the spectrum (continuum and \ion{Fe}{II} lines) in the visual, roughly 400 to 800 nm, for a model with $T_\mathrm{eff} = 7000$~K, $\log g=4.0$, and [Fe/H]$=0$.  The upper plot shows the emergent flux at the stellar surface $\mathcal{F}_\lambda$.  The blue line shows \Hm\ treated in non-LTE ($\mathcal{F}_\lambda$), and the dotted red line shows it in LTE ($\mathcal{F}_\lambda^\mathrm{LTE}$).  The lower panel shows the ratio $\mathcal{F}_\lambda / \mathcal{F}_\lambda^\mathrm{LTE}$. }  \label{fig:7000_cont}
\end{figure}

An example of the flux spectrum calculated as described in Sect.~\ref{sect:calc} for a model with T$_\mathrm{eff} = 7000$~K, $\log g=4.0$, and [Fe/H]$=0$ is shown in Fig.~\ref{fig:7000_cont}.  The general behaviour is that the increased abundance of \Hm\ around optical depth unity increases the opacity, and thus, the continuum is depressed, by about 1\% here.

\begin{figure}
\centering
\includegraphics[trim=30 80 70 120,clip,width=0.49\textwidth]{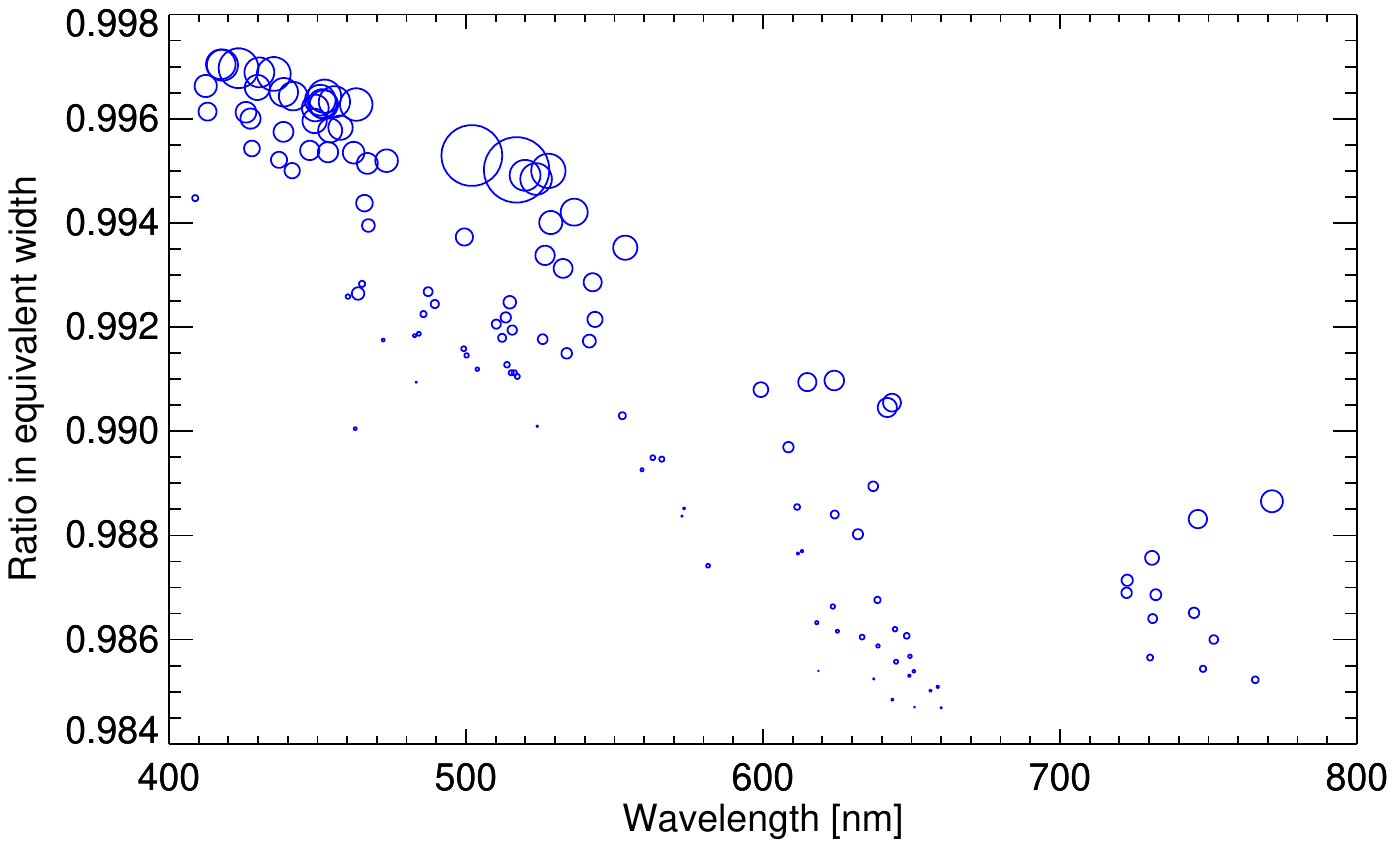}
\caption{Effect on the equivalent widths of Fe~II lines in the visual, roughly 400 to 800 nm, for a model with $T_\mathrm{eff} = 7000$~K, $\log g=4.0$, and [Fe/H]$=0$.  Each circle shows the ratio of the equivalent width for \Hm\ treated in non-LTE ($W_\lambda$) to that in LTE ($W_\lambda^\mathrm{LTE}$), i.e. $W_\lambda / W_\lambda^\mathrm{LTE}$.  The circle diameter is proportional to the equivalent width. }\label{fig:7000_lines}
\end{figure}

Figure~\ref{fig:7000_lines} shows the relative changes in the spectral line equivalent widths for this spectrum. We first note the strong trend with wavelength: The relative change increases with increasing wavelength. Secondly, at any given wavelength, strong lines are least affected in relative terms, while weak lines are more affected.  

The basic character of these effects, direction and rough magnitude, can be explained as opacity effects via the Milne-Eddington model (see, e.g. \citet[][Sect.~10.2]{mihalas_stellar_1978}, \citet[][Sect.~9.1.2]{ruttenRadiativeTransferStellar2003}, \citet[][Sect.~17.5]{hubeny_theory_2014}).  This model assumes that line shape parameters and the ratio of the line and continuum extinction, $\eta_\lambda = \alpha_\lambda^l / \alpha_\lambda^c$, are constant throughout the atmosphere.  The model assumes LTE and that the Planck function is a linear function with the continuum optical depth $B_\lambda = B_0 + B_1 \tau_\lambda^c$, corresponding to a temperature structure $T(\tau_\lambda^c)$, where $\tau_\lambda^c$ is the continuum optical depth in LTE.  The model can now be adjusted in a way analogous to the calculations presented in Sect.~\ref{sect:calc}; that is, the temperature structure is held fixed, while the continuum extinction is adjusted for the non-LTE effect on \Hm\ with departure coefficient $b_-$, also constant with depth.  Then the continuum optical depth in the non-LTE case is $b_- \tau_\lambda^{c}$, and thus $B_\lambda = B_0 + \frac{B_1}{b_-} \tau_\lambda^{c}$. 

In this model, the flux is then given by 
\begin{equation}
F_\lambda = \mathcal{F}_\lambda/\pi = B_0 + \frac{2}{3} B_1 \frac{1}{b_-+\eta_\lambda},
\end{equation}
where $F_\lambda$ is the so-called astrophysical flux, while $\mathcal{F}_\lambda$ is the emergent flux at the stellar surface \citep[see][Sect.~2.1.1]{ruttenRadiativeTransferStellar2003}.  The ratio compared to the case of LTE, $b_-=1$, has the following limiting cases:
\begin{equation}
F_\lambda / F_\lambda^\mathrm{LTE}  = 
\begin{cases} 
1,  & \text{for }   B_0 \gg B_1 \\
\frac{1+\eta_\lambda}{b_-+\eta_\lambda} &  \text{for }  B_0 \ll B_1. 
\end{cases}
\end{equation}
The first limiting case, $B_0 \gg B_1$ is trivial, an isothermal atmosphere.  For the second limiting case, closer to reality,  $B_0 \ll B_1$, the behaviour in the deep lines ($\eta_\lambda \rightarrow \infty$) and continuum ($\eta_\lambda \rightarrow 0$) gives limiting cases
\begin{equation}
F_\lambda / F_\lambda^\mathrm{LTE}  = 
\begin{cases} 
1,  & \text{for }  \eta_\lambda \rightarrow \infty  \\
1/b_- &  \text{for }  \eta_\lambda \rightarrow 0 . 
\end{cases}
\end{equation}
Thus, the flux ratio relative to LTE varies between 1 and $1/b_-$.  In the cores of deep lines, the flux is unchanged because the change in continuum extinction has no effect.  In the continuum, the flux is changed by the factor $1/b_-$, and thus, for example if $b_->1$ in the continuum-forming region near optical depth unity, then the flux is reduced and the continuum lowered.  This behaviour is shown in Fig.~\ref{fig:7000_cont}, and the approximate magnitude of the effect can be explained.  In Fig.~\ref{fig:grid} and~\ref{fig:conv}, for T$_\mathrm{eff} = 7000$~K, $\log g=4.0$, and [Fe/H]$=0$ around optical depth unity, $b_-$ varies between 1 and about 1.02. Based on the simple model presented above, the continuum is expected to be lowered by a factor in the range 1 to $1/1.02 \approx 0.98$.   Fig.~\ref{fig:7000_cont} shows that the continuum is lower by a factor $\approx 0.99$, roughly 1\%.  

Following the analysis given in the above references, assuming that the line parameters are constant with depth, and retaining the modification of a departure coefficient for the continuum extinction, we can derive expressions for the equivalent widths of spectral lines and the ratio compared to LTE. Again, it is most instructive to consider limiting cases.  For weak lines with $\eta_0 \ll 1$, where the subscript $0$ denotes the line centre, we obtain
\begin{equation}
\frac{W_\lambda}{W_\lambda^\mathrm{LTE}}  = 
\begin{cases} 
1/(b_-^2)  & \text{for }  B_0 \gg B_1 \\
1/b_-    &  \text{for }    B_1 \gg B_0, 
\end{cases}
\end{equation} 
and for strong lines $\eta_0 \gg 1$,
\begin{equation}
\frac{W_\lambda}{W_\lambda^\mathrm{LTE}} = 
\begin{cases} 
1/(b_-^{3/2})  & \text{for }  B_0 \gg B_1 \\
1/\sqrt{b_-}    &  \text{for }    B_1 \gg B_0, 
\end{cases}
\end{equation} 
noting that in the first limiting case, the equivalent widths tend towards zero.  These results, especially those for the more realistic second limiting case of $B_1 \gg B_0$, again explain the main effects and their magnitudes as shown in Fig.~\ref{fig:7000_lines}.  In particular, the relative effects on the equivalent widths of weak lines are greater than for strong lines. In strong lines, only the continuum formation depth is changed, while for weak lines, both continuum and line formation depths are affected. The dependence on wavelength, with stronger effects on equivalent widths in the red, is explained by the fact that the \Hm\ extinction increases towards its maximum near 850 nm, leading to a higher (cooler) formation depth and to a reduction of the temperature gradient in the region where the spectrum forms, and thus a reduction in $B_1$.  

\begin{figure}
\centering
\includegraphics[trim=80 80 80 120,clip,width=0.49\textwidth]{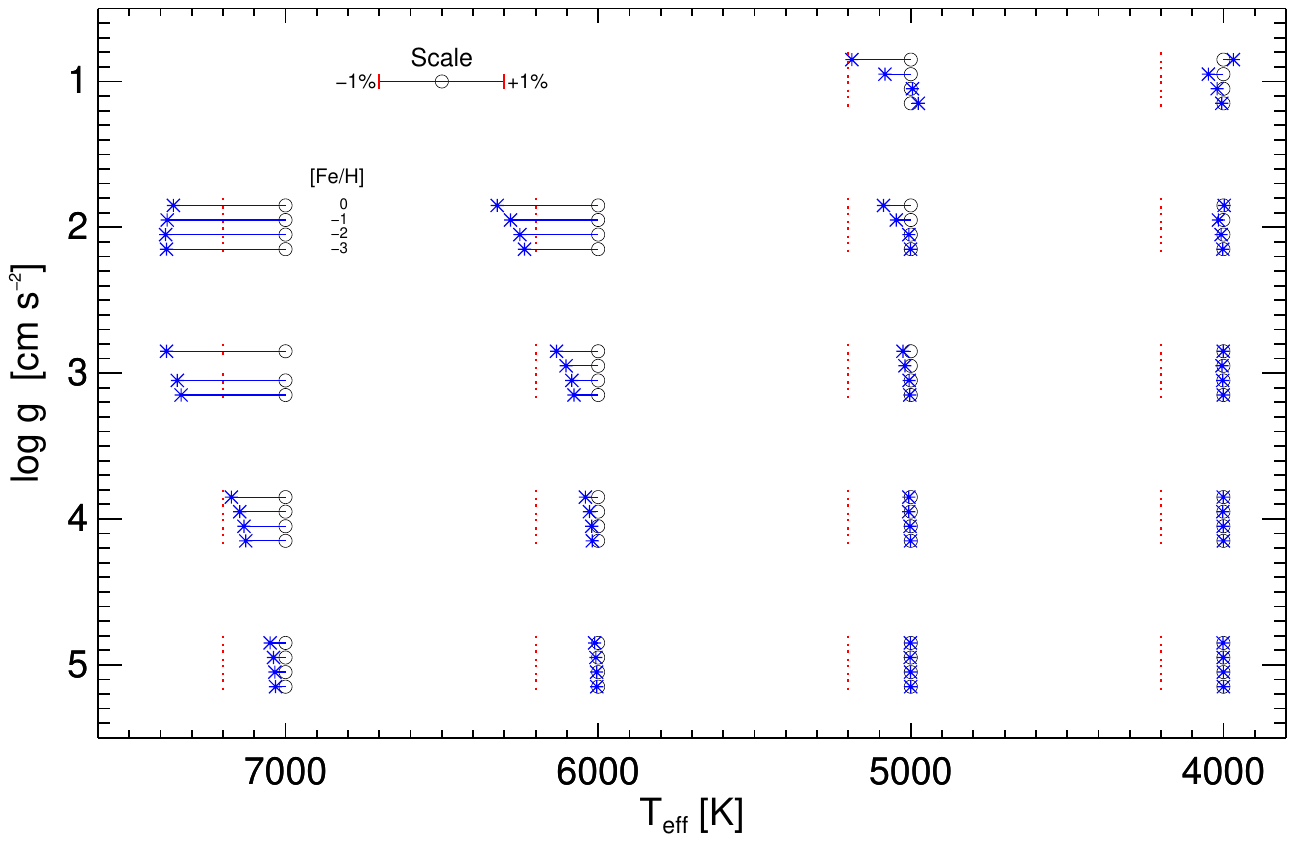}
\caption{Summary of the effects of departures from LTE for \Hm\ on the visual continuum, roughly 400 to 800 nm, across the grid of models.  For each $T_\mathrm{eff}$, $\log g$ combination, the results are shown for four [Fe/H] each slightly offset in the vertical direction such that the highest metallicity is highest in the plot.  For each case, a circle marks the position of no change, and then an asterisk displaced in the horizontal direction gives the maximum change in the continuum relative to LTE.  The scale is shown in the upper left corner, and the position of a $-1$\% change is marked in each case with a dashed red line.  For example, for the case given in Fig.~\ref{fig:7000_cont}, $T_\mathrm{eff} = 7000$~K, $\log g=4.0$, and [Fe/H]$=0$, the uppermost line in this group shows a relative change of just below $-1$\%, as shown in the lower panel of Fig.~\ref{fig:7000_cont}.} \label{fig:continuum_summary}
\end{figure}

Finally, in Fig.~\ref{fig:continuum_summary}, we attempt to summarise the effects on visual continua across our model grid.  In general, the continuum is lower by about 1\% for hotter and/or low-gravity models above the upward-sloping diagonal of the figure.  Furthermore, from solar metallicity to metal-poor atmospheres, the effects are either similar for T$_\mathrm{eff} = 7000$~K models or are reduced for cooler models.  The reasons for these behaviours were explained by LP68; in hotter stars, electrons are mainly supplied by hydrogen, and the metallicity therefore matters little.  In cooler stars, electrons are supplied by metals, and in metal-poor stars, at a given optical depth in order to retain the same total pressure, the partial pressure due to hydrogen must be higher to compensate for the lower electron pressure.  This increases the efficiency of the collision processes involving H (via \Hmol), and thus brings the system closer to LTE.  Finally, we note that for some $\log g = 1$ models, we have very small positive effects, that is, increase of the continuum, because the departure coefficient dips below unity in the region above optical depth unity (see e.g. the T$_\mathrm{eff}/\log g/ [\mathrm{Fe}/\mathrm{H}]=5000/1.0/-3.0$ model in Fig.~\ref{fig:grid}).

%-----------------------------------------------------------------
\section{Conclusions}

We have revisited the statistical equilibrium of \Hm\ in late-type stellar atmospheres, a question that has been largely untouched since the pioneering work of LP68.  Using a slightly modified version of their analytical model including H, \Hmol\, and \Hm, together with modern atomic data and a grid of 1D LTE theoretical stellar atmosphere models, we find direct non-LTE effects on populations, continua, and spectral lines of about 1--2\% in stars with higher T$_\mathrm{eff}$ and/or lower $\log g$ within the grid. The effects in models for solar parameters are smaller by a factor of 10, about 0.1--0.2\%, and for models with lower T$_\mathrm{eff}$ and/or higher $\log g$, they are practically absent. The non-LTE effects found in the spectrum-forming regions in our calculations originate from overrecombination, that is, radiative recombination of electrons with hydrogen to form \Hm\ exceeding photodetachment. 

Modern atomic data are not a source of significant differences in the results compared to LP68; the atomic data are now significantly more secure than those used by LP68, however, which places the calculations on a firm footing. In addition, the detailed data for processes on \Hmol\ resolved with vibrational and rotational states paint a more complete and complex picture of the role of \Hmol\ in the formation and equilibrium of \Hm\ than was found in LP68.  

We emphasise that this work only considered the direct effect, is only for a limited reaction network where H is assumed to be in LTE, and was furthermore conducted for 1D models.   \cite{Litesequilibriumusingcoupled1984} formulated a more complete reaction network and statistical equilibrium problem including H$_2^+$, H$^+$, and internal states of H out of LTE, but no calculations were attempted.  This should be done, ideally with the inclusion of internal states of \Hmol, and possibly also H$_3^+$.  In addition, mutual neutralisation processes of \Hm\ with metal ions, $\mathrm{H}^- + \mathrm{X}^+ \leftrightarrow \mathrm{H} + \mathrm{X}^*$ have in recent years been demonstrated to be important in non-LTE studies of neutral metal atoms, \citep[e.g.][]{Barklem2003b,Barklem2011,Belyaev2014}, and they might affect the statistical equilibrium of \Hm.  

Furthermore, the 1D modelling presented here suggests a connection to the temperature structure as modified by convection, and investigation in 3D models may be important.  Finally, modelling including both direct and indirect effects, including feedback on the atmospheric structure, should be attempted. Based on the results of \cite{shortNonLTELineBlanketedModel2005} and \cite{shapiroNLTESolarIrradiance2010}, this requires simultaneous accurate non-LTE modelling of important electron donors such as Fe, Si, and Mg.  This is a very significant challenge, even in 1D.  Consideration of all these additional factors may significantly change or remove the direct effects seen here, but it is important to understand the physics in detail, both in its own right for understanding stellar spectra and for applications in other astrophysical problems.

\begin{acknowledgements}

We thank Bengt Edvardsson for calculating \marcs{} models with different convection parameters, and doing the first calculations of $J_\nu$ used in preliminary studies, and Bengt Gustafsson for pointing out the role of convection in determining the atmospheric temperature structures deep in the atmosphere.  We thank Yohann Scribano for providing the \Hmol\ collision data.  Mar\'ila Carlos and Sema Caliskan gave comments that improved the figures.  P.S.B.\ and A.M.A.\ acknowledge support from the Swedish Research Council (VR 2020-03404 and 2020-03940). This research was supported by computational resources provided by the Australian Government through the National Computational Infrastructure (NCI) under the National Computational Merit Allocation Scheme and the ANU Merit Allocation Scheme (project y89).

\end{acknowledgements}

\bibliographystyle{aa}
\bibliography{MyLibrary_laptop}

\end{document}